\documentclass[11pt]{article}%
\usepackage{amssymb}
\usepackage[dvips]{graphicx}
\usepackage{amsmath}
\usepackage{amsfonts}%
\providecommand{\U}[1]{\protect\rule{.1in}{.1in}}

\newcommand{\BE}{\begin{equation}}
\newcommand{\EE}{\end{equation}}
\newcommand{\be}{\begin{equation}}
\newcommand{\ee}{\end{equation}}
\newcommand{\BA}{\begin{eqnarray}}
\newcommand{\EA}{\end{eqnarray}}

\newcommand{\lphfour}{(\lambda \Phi^4)_4}

\begin{document}
\begin{titlepage}
\vspace*{22mm}
\begin{center}
{\LARGE{\bf  The long-range interaction}} \\
\vspace*{2mm}
{\LARGE{\bf in massless $(\lambda \Phi^4)_4$ theory}}\\
\vspace*{20mm}
{\Large P. M. Stevenson}
\vspace*{3mm}\\
{\large T. W. Bonner Laboratory, Department of Physics and Astronomy, \\
Rice University, P.O. Box 1892, Houston, TX 77251-1892, USA}
\vspace{32mm}\\
{\bf Abstract:}
\end{center}
Does massless $\lphfour$ theory exhibit spontaneous symmetry breaking
(SSB)?
The raw 1-loop result implies that it does, but the ``RG-improved'' result
implies the opposite.  I argue that the appropriate ``low-energy effective
theory'' is a nonlocal field theory involving an attractive, long-range
interaction $\Phi^2(x)\Phi^2(y)/z^4$, where $z= \mid\! x-y \!\mid$.
RG improvement then requires running couplings for both this interaction and
the original pointlike interaction.  A crude calculation in this framework
yields SSB even after ``RG improvement'' and closely agrees with the
raw 1-loop result.
\end{titlepage}

\newpage

\section{Introduction}

Although seemingly the simplest renormalizable quantum field theory, the
4-dimensional $\lambda\Phi^{4}$ theory differs from other paradigm theories,
QED and QCD -- and, as yet, there is no experimental confirmation that we
understand the theory correctly. The issue is important because of the
relevance of $\lambda\Phi^{4} $ theory to the Higgs mechanism and to
inflationary cosmology.

One basic issue is the nature of the phase transition as one varies the
renormalized-mass $m$: Is it a second-order transition occurring at $m^{2}=0$,
or is it a first-order transition occurring at some small but positive value
of $m^{2}$? Equivalently, one can ask if the massless theory is still in the
symmetric phase or not. The dilemma goes back to Coleman and Weinberg
\cite{CW} who found that in massless $\lambda\Phi^{4}$ the 1-loop effective
potential (1LEP) predicts spontaneous symmetry breaking (SSB), but that
``renormalization-group improvement'' (RGI) contradicts that result. (By
contrast, in Scalar Electrodynamics they found that RGI \textit{confirmed} the
1-loop prediction of SSB.) It is the contradiction between 1LEP and RGI in
$\lambda\Phi^{4}$ theory that is the main concern of this paper.

\section{A particle-language argument}

A very physical argument for a first-order transition was given in Ref.
\cite{mech}. Consider the $\lambda\Phi^{4}$ theory with a small, positive
renormalized mass $m$ and focus on the particles (\textquotedblleft
phions\textquotedblright) and their interactions. The fundamental interaction
is the pointlike repulsive interaction (Fig. 1(a)). However, the `fish'
diagram (Fig. 1(b)) induces a long-range attractive interaction, proportional
to $-1/r^{3}$ if we ignore the phion mass \cite{fein,mech}. Consider a large
number $\mathsf{N}$ of phions contained in a large box of volume $\mathsf{V}$
with periodic boundary conditions. What is the lowest energy of this system?
Naturally, to minimize the kinetic energy we would put almost all the phions
into the $\mathbf{k}=0$ mode. Thus, if there were no interactions, the energy
would just be the sum of the rest-energies, $\mathsf{N}m$. Two-body
interactions add a term given by the number of pairs (${\scriptstyle{\frac
{1}{2}}}\mathsf{N}(\mathsf{N}-1)\approx{\scriptstyle{\frac{1}{2}}}%
\mathsf{N}^{2}$) multiplied by the average interparticle potential energy
$\bar{u}$, giving a total energy
\begin{equation}
E_{\mathrm{tot}}=\mathsf{N}m+{\scriptstyle{\frac{1}{2}}}\mathsf{N}^{2}\bar{u}.
\end{equation}
Effects from three-body or multi-body interactions will be negligible provided
that the phion gas is dilute.

%
\begin{figure}[ptb]
\begin{center}
\includegraphics[height=1.1467in,width=4.2134in]{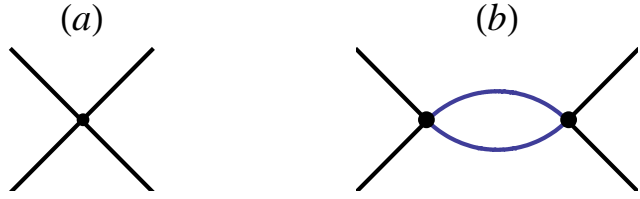}
\end{center}
\caption{(a) The fundamental interaction. \ (b) The \textquotedblleft
fish\textquotedblright\ diagram, which induces a long-range interaction.}%
\end{figure}

If the interparticle potential is $U(r)$ and we assume that the particles are
uniformly distributed over the box, which is true if almost all the particles
are in the $\mathbf{k}=0$ mode, then
\begin{equation}
\label{ubar}\bar{u} \sim{\frac{{1}}{{\mathsf{V}}}}\int\! d^{3}r \, U(r).
\end{equation}
Hence, we obtain an energy density
\begin{equation}
\label{Eest}\mathcal{E} \equiv E_{\mathrm{tot}} /\mathsf{V} = n m +
{\scriptstyle{\frac{1}{2}}} n^{2} \int\! d^{3} r \, U(r),
\end{equation}
where $n \equiv\mathsf{N}/\mathsf{V}$. The interparticle potential $U(r)$
consists of the pointlike repulsion term plus the $-1/r^{3}$ term. The former
integrates to some constant, but the latter requires cutoffs at both small and
large $r$, giving
\begin{equation}
\mathcal{E} = n m + C_{1} n^{2} - C_{2} n^{2} \ln(r_{\mathrm{max}}/r_{0}),
\end{equation}
where $C_{1}, C_{2}$ are some positive constants. The ultraviolet divergence
is not unexpected and $1/r_{0}$ is naturally identified with the ultraviolet
cutoff needed in the field theory. The infrared divergence arises, of course,
from neglecting the mass of the exchanged phions in the `fish' diagram;
actually the $-1/r^{3}$ potential is exponentially suppressed at distances $r
> 1/(2m)$. However, when $m$ is very small another consideration is more
important; namely, that the long-distance attraction between two phions will
be ``screened'' by other phions that interpose themselves. This consideration
implies an effective $r_{\mathrm{max}}$ that depends on the density $n$.

[The virtual phions being exchanged undergo multiple scatterings with the
background $\mathbf{{k}=0}$ phions (See Fig. 2). Each collision corresponds to
a mass insertion and the resulting geometric series changes the mass to
$M^{2}=m^{2}+\lambda n/m$. The mass $M$ represents the mass of a quasiparticle
excitation of the phion condensate; i.e., it is the Higgs boson mass. In the
small-$m$ regime of interest $M\approx\lambda n/m$ and so $r_{\mathrm{max}%
}=1/(2M)\propto1/\sqrt{n}$.]

%
\begin{figure}[ptb]
\begin{center}
\includegraphics[height=1.7236in,width=3.2128in]{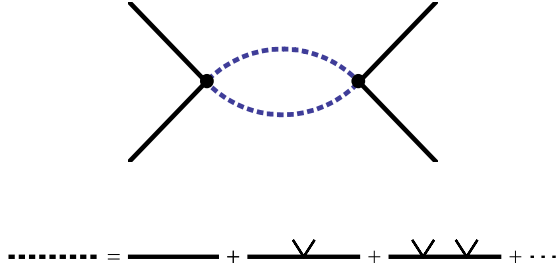}
\end{center}
\caption{Screening of the long-range interaction due to multiple collisions
with background phions.}%
\end{figure}

\bigskip

Hence, $\mathcal{E}$ is given by a sum of $n$, $n^{2}$ and $n^{2}\ln n$ terms
which represent, respectively, the rest-energy cost, the repulsion energy
cost, and the energy gain from the long-range attraction. If the rest-mass $m$
is small enough, then the $n^{2}\ln n$ term's negative contribution will
result in an energy density whose global minimum is not at $n=0$ but at some
specific, non-zero density $n_{v}$. That is, even though the `empty' state is
locally stable, it can decay by spontaneously generating particles so as to
fill the box with a dilute condensate of density $n_{v}$. This condensate is
the the SSB vacuum. In field-theory terms the particle density $n$ is
proportional to the classical field squared, $n={\scriptstyle{\frac{1}{2}}%
}m\phi^{2}$, and the energy density as a function of $n$ is the
field-theoretic \textit{effective potential}: $\mathcal{E}(n)\equiv
V_{\mathrm{eff}}(\phi)$. This argument, then, reproduces the form of the 1LEP,
with $\phi^{2}$, $\phi^{4}$, and $\phi^{4}\ln\phi^{2}$ terms.

\section{Implications for RG improvement}

An important moral of the above argument is that in nearly-massless
$\lambda\Phi^{4}$ theory there are really \textit{two} interactions; a
short-range repulsion and a long-range attraction, the latter induced by the
former. Successful use of RG methods requires that we first identify all
appropriate interactions (``relevant'' and ``marginal'') and then consider how
the strengths of those interactions change with scale. The suggestion here is
that we need to consider \textit{two} running couplings.

The usual RGI procedure in massless $\lambda\Phi^{4}$ theory considers only
the pointlike $\lambda\phi^{4}$ interaction. The main effect of loop diagrams,
it asserts, is just to renormalize $\lambda$, resulting in a running coupling
$\lambda_{R}(\mu)$ that runs with renormalization scale $\mu$. The effective
potential, to lowest order, is then the classical potential with $\lambda$
replaced by $\lambda_{R}(\phi)$:
\begin{equation}
V_{\mathrm{eff}}^{\mathrm{RGI}} = \frac{\lambda_{R}(\phi)}{4!} \phi^{4}.
\end{equation}
Since, to leading order, $\lambda_{R}(\phi)$ is proportional to $1/\ln\phi
^{2}$, this RGI effective potential does not show SSB. Higher-order versions
of this RGI procedure do not change this qualitative result.

However, when we contemplate the ``fish'' diagram (Fig. 1(b)), we see that,
while its short-distance part containing the ultraviolet divergence does
indeed contribute to renormalizing $\lambda$, its long-distance part is
actually creating a physically distinct interaction, the long-range attraction
(a $-1/r^{3}$ interparticle potential in the particle language of the previous
section). If we now switch to a manifestly covariant, Euclideanized
functional-integral formalism, the ``fish'' diagram will generate a term in
the effective action proportional to
\begin{equation}
\int\! d^{4} x \int\! d^{4} y \, \Phi^{2}(x) (\mathcal{G}_{xy})^{2} \Phi
^{2}(y),
\end{equation}
where $\mathcal{G}_{xy}$ is the $x$-space propagator, proportional to
$1/\!\mid\! x-y \! \mid^{2}$ if we neglect the phion mass. This term is
inherently non-local. While a fundamental Lagrangian must be local, for
reasons of causality, an effective Lagrangian can, and in general does,
contain non-local terms. In theories without infrared sensitivity one may make
a derivative expansion of the non-local terms as a series of local terms, but
that option is not possible here.\footnote{If the interaction kernel
$1/\!\mid\! x-y \! \mid^{4}$ had fallen off with any higher power of distance
then it could have been treated in this manner. In particle language, if the
interparticle potential had fallen off faster than $-1/r^{3}$ then it would
produce a finite scattering length and could be approximated by a local
pseudopotential, as in atomic physics.}

This paper does not attempt to present a comprehensive formalism embodying
these ideas. Instead, for maximal simplicity, we set aside some important
issues (precisely how the non-local term is generated and how to avoid double
counting) and simply start directly with a non-local action.

\section{The nonlocal model}

Consider then a nonlocal effective theory described by the following Euclidean
action:
\begin{equation}
\label{SNL}S= \int\! d^{4} \, x \, \frac{1}{2} \partial_{\mu} \Phi\,
\partial^{\mu} \Phi+ \int\! d^{4} x \! \int\! d^{4} y \,\, \Phi^{2}(x) U(x-y)
\Phi^{2}(y),
\end{equation}
where the interaction kernel $U(z)$ is a function only of the Euclidean
distance $z= \sqrt{z_{\mu}z^{\mu}}$. It consists of a short-range repulsive
core and a long-range attractive tail:
\begin{equation}
U=U_{\mathrm{core}} + U_{\mathrm{tail}},
\end{equation}
where
\begin{equation}
U_{\mathrm{core}}(z) = \frac{\lambda}{4!} f(z;z_{0}) \quad\quad\quad
{\scriptstyle{(0<z<z_{0})}},
\end{equation}
and
\begin{equation}
U_{\mathrm{tail}}(z) = - \frac{\zeta}{z^{4}} \quad\quad\quad
{\scriptstyle{(z_{0}<z<z_{\mathrm{max}})}}.
\end{equation}
Here, $\lambda$ and $\zeta$ are positive parameters and $\int_{0}^{z_{0}}
d^{4} z f(z;z_{0}) \equiv1$ so that $f(z;z_{0})$ approaches $\delta^{(4)}(z)$
as $z_{0} \to0$. It is convenient to define a positive, dimensionless coupling
$\eta$ for the long-range interaction by:
\begin{equation}
\label{etadef}\frac{\eta}{4!} \equiv- \int\! d^{4} z \, U_{\mathrm{tail}}(z) =
2 \pi^{2} \zeta\ln\left(  \frac{z_{\mathrm{max}}}{z_{0}} \right)  ,
\end{equation}
so that
\begin{equation}
\lambda_{\mathrm{tot}} \equiv\lambda- \eta
\end{equation}
represents the ``total'' interaction strength, in the sense that
\begin{equation}
\frac{\lambda_{\mathrm{tot}}}{4!} = \int d^{4} z \left(  U_{\mathrm{core}}(z)
+ U_{\mathrm{tail}}(z) \right)  .
\end{equation}

The theory is also to be regulated by an ultraviolet cutoff $\Lambda$ by
excluding all momentum modes with $p>\Lambda$. Note that, at this stage,
$\Lambda, z_{0}, z_{\mathrm{max}}$ and $\lambda, \eta$ are all independent,
free parameters of the model theory. Later on we shall be interested in taking
limits where $\Lambda\to\infty$, $z_{0} \to0$, and $z_{\mathrm{max}} \to
\infty$. (Note that $z_{\mathrm{max}}$ can be thought of as $1/(2m)$, where
$m$ is the phion mass, which is otherwise neglected.)

The aim now is to compute the effective potential of this model theory to
second order in a double perturbation expansion in powers of the renormalized
$\lambda$ and $\eta$ couplings. This second-order result can be obtained from
the 1-loop calculation. To begin, one re-writes the action (\ref{SNL}) after
shifting the field, $\Phi(x)=\phi+h(x)$:
\begin{equation}
\label{Ssh}S= \mathcal{V} \, \frac{\lambda_{\mathrm{tot}}}{4!} \phi^{4} +
\frac{1}{2} \int\! d^{4} x \! \int\! d^{4} y \, h(x) \mathcal{G}^{-1}_{xy}
h(y) + (h^{3}, h^{4} {\mbox{\rm terms}}),
\end{equation}
where $\mathcal{V}$ is the infinite spacetime volume factor $\int\! d^{4} x$,
and the inverse propagator is
\begin{equation}
\mathcal{G}^{-1}_{xy} = \left(  - \partial^{2} + {\scriptstyle{\frac{1}{6}}}
\lambda_{\mathrm{tot}} \phi^{2} \right)  \delta^{(4)}(x-y) + 8 \phi^{2}
U(x-y).
\end{equation}
Fourier transforming to momentum space leads to
\begin{equation}
\tilde{\mathcal{G}}^{-1}(p) = p^{2} + {\scriptstyle{\frac{1}{6}}}
\lambda_{\mathrm{tot}} \phi^{2} + g_{c}(p) + g_{t}(p),
\end{equation}
with
\begin{equation}
g_{c}(p) = 8 \phi^{2} \int\! d^{4} z \, e^{i p.z} \, U_{\mathrm{core}}(z)
\quad\quad( \to{\scriptstyle{\frac{1}{3}}} \lambda\phi^{2} \quad{\mbox{\rm
for}} \quad z_{0} \to0),
\end{equation}
\begin{equation}
g_{t}(p) = 8 \phi^{2} \int\! d^{4} z \, e^{i p.z} \, U_{\mathrm{tail}}(z) = -
{\scriptstyle{\frac{1}{3}}} \eta\phi^{2} \left(  1 - \frac{( S(p
z_{\mathrm{max}})-S(p z_{0}) )}{\ln(z_{\mathrm{max}}/z_{0})} \right)  ,
\end{equation}
where
\begin{equation}
S(u) = -2 \int_{0}^{u} \frac{dv}{v^{2}}(J_{1}(v)-{\scriptstyle{\frac{1}{2}}}
v)
\end{equation}
and $J_{1}(v)$ is a Bessel function. $S(u)$ can be expressed in terms of a
hypergeometric function;
\begin{equation}
S(u) = \frac{1}{16} u^{2} F_{2,2,3}^{1,1}(-{\scriptstyle{\frac{1}{4}}} u^{2})
\end{equation}
and its small- and large-argument behaviours are
\begin{equation}
S(u) \sim\frac{1}{16} u^{2} \quad\quad\quad(u \to0),
\end{equation}
\begin{equation}
S(u) \sim\ln(u/\sigma)-{\scriptstyle{\frac{1}{2}}} \quad\quad(u \to\infty),
\end{equation}
with $\sigma= 2 e^{-\gamma}=1.12 \ldots$, where $\gamma$ is the Euler
constant. (In fact, the above equations provide a good qualitative description
of $S(u)$ for $u \lesssim2.8$ and $u \gtrsim2.8$, respectively.) Note that
$g_{t}(p)$ is $-{\scriptstyle{\frac{1}{3}}} \eta\phi^{2}$ at $p=0$ and tends
to zero as $p \to\infty$.

\section{1LEP for the nonlocal model}

A full calculation of the 1-loop effective potential in this nonlocal model,
valid for a wide range of the parameters $\Lambda,z_{0},z_{\mathrm{max}}$, has
been carried out in collaboration with S.~A.~Hassan \cite{Asif}. Here, for
maximal simplicity, we consider a limit in which $z_{\mathrm{max}}%
\rightarrow\infty$ (i.e., $m\rightarrow0$) \textit{before} taking the
$\Lambda\rightarrow\infty$ and $z_{0}\rightarrow0$ limits. (The full
calculation reproduces our results below provided $\mid\!\ln(mz_{0})\!\mid
\gg\mid\!\ln(M_{v}z_{0})\!\mid$.) \ Then, for all momenta in the range
$1/z_{\mathrm{max}}\ll p\ll1/z_{0}$ (which will become essentially all finite,
non-zero momenta) the function $g_{t}(p)$ becomes negligible, of order
$\eta\phi^{2}/\ln(z_{\mathrm{max}}/z_{0})$. However, for $p=0$, we have, as
always, $g_{t}(p=0)=-{\scriptstyle{\frac{1}{3}}}\eta\phi^{2}$. Thus, the bare
propagator has a discontinuity at zero momentum:
\begin{equation}
\tilde{\mathcal{G}}(p)=\left\{
\begin{array}
[c]{ll}%
{\scriptstyle{\frac{1}{2}}}\lambda_{\mathrm{tot}}\phi^{2} & \quad\quad p=0\\
p^{2}+M^{2} & \quad\quad p\neq0
\end{array}
\right.  \label{invprop1}%
\end{equation}
with
\begin{equation}
M^{2}\equiv M^{2}(\phi)=\left(  {\scriptstyle{\frac{1}{6}}}\lambda
_{\mathrm{tot}}+{\scriptstyle{\frac{1}{3}}}\lambda\right)  \phi^{2}%
={\scriptstyle{\frac{1}{2}}}(\lambda-{\scriptstyle{\frac{1}{3}}}\eta)\phi^{2}.
\end{equation}
In this limit the 1-loop effective potential can then be obtained immediately
as
\begin{equation}
V_{\mathrm{eff}}^{\mathrm{1-loop}}(\phi)=\frac{\lambda_{\mathrm{tot}}}{4!}%
\phi^{4}+\frac{1}{2}\int\!\frac{d^{4}p}{(2\pi)^{4}}\ln(p^{2}+M^{2}).
\end{equation}
After subtracting an infinite constant and removing all $\phi^{2}$ terms,
corresponding to a mass renormalization condition $\left.  d^{2}%
V_{\mathrm{eff}}/d\phi^{2}\right\vert _{\phi=0}=0$, this becomes
\begin{equation}
V_{\mathrm{eff}}^{\mathrm{1-loop}}(\phi)=\frac{\phi^{4}}{4!}\left(
(\lambda-\eta)-\frac{3}{16\pi^{2}}(\lambda-{\scriptstyle{\frac{1}{3}}}%
\eta)^{2}\left(  \ln\frac{\Lambda}{M}+\frac{1}{4}\right)  \right)  .
\end{equation}

Obviously, this expression reduces to the usual one when $\eta=0$, but note
that now the first term involves $\lambda-\eta$ while the second involves a
different combination $\lambda-{\scriptstyle{\frac{1}{3}}} \eta$. One may now
proceed to renormalize perturbatively by defining renormalized couplings
$\lambda_{R}, \eta_{R}$ to absorb the ultraviolet divergence. At first sight,
it seems that one can only determine the divergence present in the
\textit{combination} $\lambda-\eta$. However, two simple arguments lead to a
unique solution: (i) the short-range interaction, $\lambda$, should only be
renormalized by short-range interactions --- since a short-range interaction
modified by a long-range interaction would cease to be short range; (ii)
Although the short-range interaction, acting twice, induces a long-range
interaction, there is no $\ln\Lambda$ divergence involved in the calculation.
Argument (i) implies that $\lambda_{R}$ involves only $\lambda$, while
argument (ii) implies that $\eta_{R}$ involves $\ln\Lambda$ divergences only
in the $\eta\lambda$ and $\eta^{2}$ terms, not in the $\lambda^{2}$ term.
Thus, the renormalization is accomplished by
\begin{equation}
\label{renlam}\lambda_{R}(\mu) = \lambda- \lambda^{2} \left(  \frac{3}{16
\pi^{2}} \right)  \left(  \ln\frac{\Lambda}{\mu} + a_{1} \right)  + \ldots,
\end{equation}
\begin{equation}
\label{reneta}\eta_{R}(\mu) = \eta+ \eta^{2} \left(  \frac{1}{48 \pi^{2}}
\right)  \left(  \ln\frac{\Lambda}{\mu} + a_{2} \right)  + \eta\lambda\left(
-\frac{1}{8 \pi^{2}} \right)  \left(  \ln\frac{\Lambda}{\mu} + a_{3} \right)
+ a_{4} \lambda^{2} + \ldots,
\end{equation}
where $\mu$ is some arbitrary renormalization scale and $a_{1}, a_{2}, a_{3}$
are arbitrary, renormalization-scheme-dependent finite constants. The $a_{4}$
term is a ``calculable coefficient'':
\begin{equation}
a_{4}= \frac{3}{2 \pi^{2}} \lambda^{2} \ln(z_{\mathrm{max}}/z_{0})
\end{equation}
except that it involves an infinite $\ln(z_{\mathrm{max}}/z_{0})$ factor from
converting $\zeta$ to $\eta$ (see Eq. (\ref{etadef})). However, all that
matters here is that $a_{4}$ has no $\Lambda$ or $\mu$ dependence and so will
not affect any beta functions. In fact, one may argue that this term should
simply be subtracted to avoid double counting of the long-range interaction.

[Eqs. (\ref{renlam}), (\ref{reneta}) can also be deduced from the 1-loop
diagrams for the 4-point function in this nonlocal theory. The correct
combinatorial factors, reported in Fig. 3, can be found from a
straightforward, if tedious, calculation of the fourth functional derivative
of the generating functional $Z[j]$ associated with the nonlocal action
(\ref{SNL}).]

%
\begin{figure}[ptb]
\begin{center}
\includegraphics[height=3.0268in,width=3.6339in]{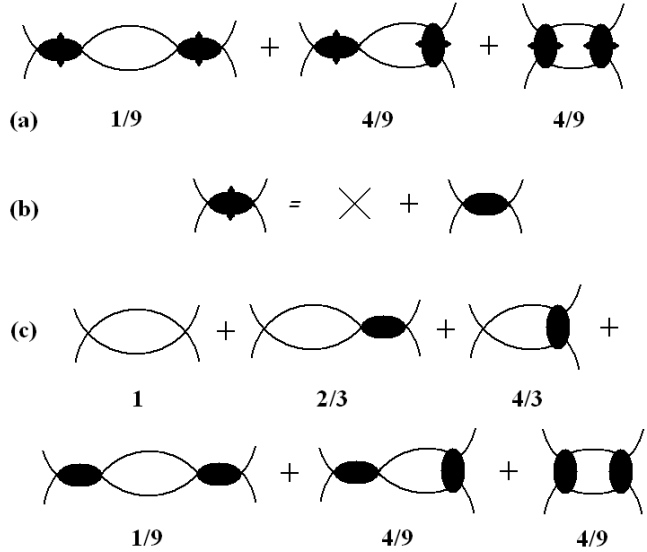}
\end{center}
\caption{(a) With a nonlocal interaction $U$ the fish diagram generalizes to
three distinct diagrams with combinatoric weights as shown. Substituting
$U=U_{\mathrm{core}}+U_{\mathrm{tail}}$, as shown in (b), leads to the weights
shown in (c). The second and fourth diagrams in (c) produce the $-\eta\lambda$
and $\eta^{2}$ divergences in Eq. (\ref{reneta}).}
\end{figure}

\bigskip

Substituting for the bare parameters (using the inverse of Eqs. (\ref{renlam}%
), (\ref{reneta})) and re-expanding in powers of the renormalized couplings,
dropping terms cubic or higher, leads to the renormalized perturbative result.
It consists solely of $\phi^{4}$ and $\phi^{4}\ln\phi$ terms, so that if we
define $v$ as the location of the minimum (i.e., the vacuum expectation of
$\phi$) we may write the result as
\begin{equation}
V_{\mathrm{eff}}^{\mathrm{pert}}=\frac{M_{R}^{4}}{64\pi^{2}}\left(  \ln
\frac{\phi^{2}}{v^{2}}-\frac{1}{2}\right)  , \label{Vpert}%
\end{equation}
with
\begin{equation}
M_{R}^{2}=\frac{1}{2}\left(  \lambda_{R}-{\scriptstyle{\frac{1}{3}}}\eta
_{R}\right)  \phi^{2}.
\end{equation}
(Note that the constants $a_{1},\ldots,a_{4}$ do not appear in this result;
they are subsumed in $v$.)

Thus, in the non-local model, just as in the usual massless $(\lambda\Phi
^{4})_{4}$ theory, the one-loop result shows SSB. The question, though, is
whether or not this result persists after ``RG improvement.''

\section{``Improved'' RG improvement}

We next consider the RG-improvement procedure in the nonlocal model. The RG
beta functions follow immediately from Eqs. (\ref{renlam}), (\ref{reneta})
above:
\begin{equation}
\label{betalam}\beta_{\lambda} (\lambda_{R}) \equiv\mu\frac{\partial
\lambda_{R}}{\partial\mu} = \frac{3}{16 \pi^{2}} \lambda_{R}^{2} + \ldots,
\end{equation}
\begin{equation}
\label{betaeta}\beta_{\eta} (\eta_{R},\lambda_{R}) \equiv\mu\frac{\partial
\eta_{R}}{\partial\mu} = -\frac{1}{48 \pi^{2}} \eta_{R}^{2} + \frac{1}{8
\pi^{2}} \eta_{R} \lambda_{R} + \ldots.
\end{equation}
Integrating (\ref{betalam}) to leading order gives, as usual,
\begin{equation}
\lambda_{R}(\mu) = \frac{16 \pi^{2}}{3} \frac{1}{\ln\mu_{0}/\mu},
\end{equation}
where $\mu_{0}$ is the Landau-pole scale. Substituting into (\ref{betaeta})
yields the differential equation
\begin{equation}
\mu\frac{\partial\eta_{R}}{\partial\mu} - \frac{2}{3} \frac{\eta_{R}}{\ln
\mu_{0}/\mu} = - \frac{1}{48 \pi^{2}} \eta_{R}^{2}.
\end{equation}
Multiplying by the integrating factor $(\ln\mu_{0}/\mu)^{2/3}$, the left-hand
side becomes a total derivative, yielding the solution
\begin{equation}
\eta_{R}(\mu) = \frac{16 \pi^{2}}{u^{2} (u_{c}-u)},
\end{equation}
where
\begin{equation}
u \equiv(\ln\mu_{0}/\mu)^{1/3}%
\end{equation}
and $u_{c}$ is a constant of integration. The $\eta_{R}$ coupling thus has
singularities at both the Landau scale $\mu_{0}$ ($u=0$) and at a scale
$\mu_{c}$ ($u=u_{c}$). To have $\eta_{R}$ positive and a region where both
$\lambda_{R}$ and $\eta_{R}$ are small requires $\mu_{c} \ll\mu_{0}$. That
means that $u_{c} \equiv(\ln\mu_{0}/\mu_{c} )^{1/3}$ should be a large number.
Figure 4 shows the running couplings as a function of $u$.

%
\begin{figure}[ptb]
\begin{center}
\includegraphics[height=2.5806in,width=3.5993in]{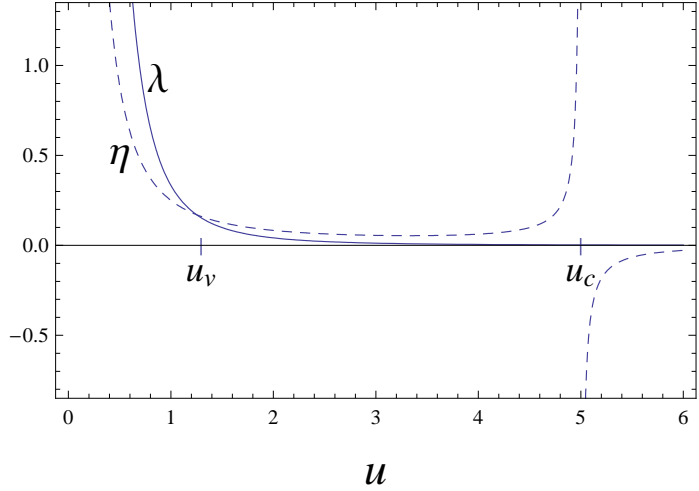}
\end{center}
\caption{Running couplings (divided by $16\pi^{2}$) as functions of
$u\equiv(\ln(\mu_{0}/\mu))^{1/3}$. (Shown for $u_{c}=5$.) }%
\end{figure}

\bigskip

The \textquotedblleft RG improved\textquotedblright\ effective potential, to
leading order, is obtained by taking the classical potential (the first term
of Eq. (\ref{Ssh}) with $\mathcal{V}$ divided out) and replacing the bare
coupling with a running coupling: i.e.,
\begin{equation}
V_{\mathrm{eff}}^{\mathrm{RGI}}(\phi)=\frac{\lambda_{\mathrm{tot}}^{R}(\phi
)}{4!}\phi^{4},
\end{equation}
where
\begin{equation}
\lambda_{\mathrm{tot}}^{R}(\mu)=\lambda_{R}(\mu)-\eta_{R}(\mu)=\frac{16\pi
^{2}}{3}\frac{(u_{c}-4u)}{u^{3}(u_{c}-u)},
\end{equation}
and the renormalization scale $\mu$ is chosen to be $\phi$. Taking the
derivative of $V_{\mathrm{eff}}^{\mathrm{RGI}}(\phi)$ (noting that $\mu
du/d\mu=-1/(3u^{2})$) yields
\begin{equation}
\frac{dV_{\mathrm{eff}}^{\mathrm{RGI}}}{d\phi}=\frac{2\pi^{2}}{9}\phi^{3}%
\frac{\left[  4u^{3}(4u-u_{c})(u-u_{c})+(2u-u_{c})^{2}\right]  }{u^{6}%
(u_{c}-u)^{2}}.
\end{equation}
One sees that the factor in square brackets has a non-trivial root close to
$u={\scriptstyle{\frac{1}{4}}}u_{c}$:
\begin{equation}
u_{v}=\frac{1}{4}u_{c}\left(  1+\frac{16}{3}\frac{1}{u_{c}}^{3}+\mathcal{O}%
\left(  \frac{1}{u_{c}^{6}}\right)  \right)  .
\end{equation}
This means that $V_{\mathrm{eff}}^{\mathrm{RGI}}(\phi)$ has an SSB minimum at
$\phi=v$, where $(\ln\mu_{0}/v)^{1/3}=u_{v}$, that is $v=\mu_{0}\exp
(-u_{v}^{3})$. The form of $V_{\mathrm{eff}}^{\mathrm{RGI}}(\phi)$ is shown in
Fig. 5. For large $u_{c}$ the RGI effective potential is very close to the
\textquotedblleft unimproved\textquotedblright\ perturbative result of Eq.
(\ref{Vpert}), except for a weak pole at very small $\phi$, corresponding to
the infrared pole in $\eta_{R}$ at $u=u_{c}$. It seems reasonable to dismiss
this infrared pole as an artifact of the leading-order approximation, perhaps
signalling some complex physics in the far infrared. Higher terms in the
$\eta$ beta function might well remove the pole, giving a \textquotedblleft
freezing\textquotedblright\ of $\eta$ in the infrared.

%
\begin{figure}[ptb]
\begin{center}
\includegraphics[height=2.738in,width=3.9634in]{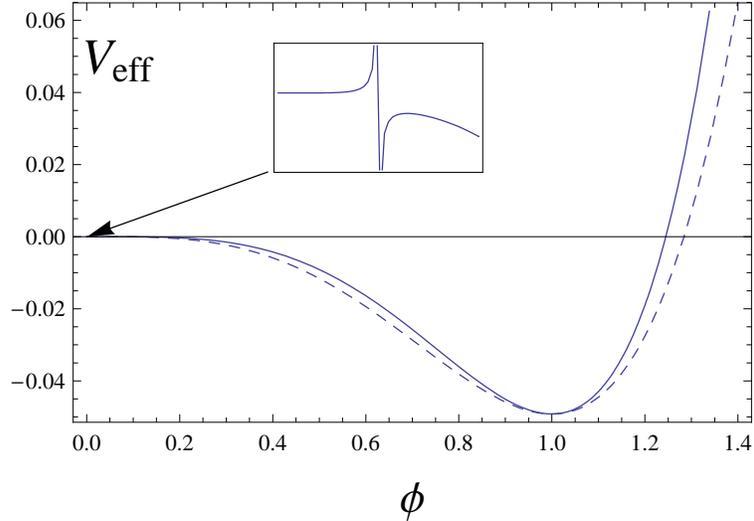}
\end{center}
\caption{RGI effective potential (solid line) compared with the raw
perturbative result (dashed curve). Units such that $v=1$. The RGI result has
a weak pole at a tiny value of $\phi$. In this example $u_{c}=5$. For larger
$u_{c}$ the two curves become almost indistinguishable and the pole, here at
$\phi\approx4\times10^{-54}$, retreats to exponentially smaller values of
$\phi$.}%
\end{figure}

\bigskip

The important point is that one now has essential agreement between the
perturbative and RGI results, provided $u_{c}$ is large (which is necessary
for the renormalized couplings to be small at the scale $v$). In the
conventional treatment the RGI effective potential cannot show SSB because the
running coupling $\lambda_{R}(\phi)$ must remain positive. Here, although
$\lambda_{R}$ remains positive, the RGI effective potential is governed by
$\lambda_{\mathrm{tot}}^{R}=\lambda_{R}-\eta_{R}$ which can and does go
negative. The SSB vacuum is very close to $u={\scriptstyle{\frac{1}{4}}}u_{c}$
where $\lambda_{R}$ and $\eta_{R}$ cancel. Hence, the inverse propagator
(\ref{invprop1}) nearly vanishes at $p=0$, in accord with \cite{cons}.

\section{Conclusions}

The massless $(\lambda\Phi^{4})_{4}$ theory has two physically distinct
interactions -- the fundamental short-range repulsion and the induced
long-range attraction. Conventional RGI ignores the latter. RGI including both
interactions leads to SSB and closely agrees with the raw 1-loop result.

\vspace*{3mm}

\textbf{Acknowledgments:} Discussions with S.~A.~Hassan and M.~Consoli are
gratefully acknowledged.


\end{document}